\documentclass[12pt]{article} \usepackage{hyperref} 
 
\def\be{\begin{equation}} \def\ee{\end{equation}} 
\def\bea{\begin{eqnarray}} \def\eea{\end{eqnarray}} 

\makeatletter \def\section{\@startsection {section}{1}{\z@}{-3.5ex 
plus -1ex minus -.2ex}{2.3ex plus .2ex}{\large\bf}} 
\def\subsection{\@startsection{subsection}{2}{\z@}{-3.25ex plus -1ex 
minus -.2ex}{1.5ex plus .2ex}{\normalsize\bf}} 
 
\usepackage{amssymb,amsmath,amsfonts,graphicx} \usepackage{epsfig} 
 
\newcommand{\id}{\hbox{1\kern-.27em l}} 
\newcommand{\sid}{\hbox{\scriptsize1\kern-.27em l}} 

\newcommand{\we}{\kern-.1em\wedge\kern-.1em} 
\newcommand{\scal}{\kern-.13em\cdot\kern-.13em}

\newcommand{\II}{I\kern-.09em I}

\newcommand{\Z}{\mathbb{Z}}

\newcommand{\spa}{\ \ ,\ \ \ \ }

\begin{document} 
\begin{titlepage} 
\rightline{NORDITA-2002/70 HE} 
 
\vskip 2.5cm \centerline{\LARGE  A note on the dual of} 
\vskip 0.5cm 
 \centerline{\LARGE ${\cal N}=1$ super Yang-Mills theory} 
 
\vskip 1.5cm

\centerline{\large M. Bertolini and P. Merlatti} 
 
\vskip .8cm 
 
\centerline{\sl NORDITA} 
 
\centerline{Blegdamsvej 17, 2100 Copenhagen \O, Denmark} 
 
\centerline{\tt teobert@nbi.dk, merlatti@nbi.dk} 
 
\vskip 2cm 
 
\begin{abstract} 
\noindent We refine the dictionary of the gauge/gravity correspondence 
realizing ${\cal N}=1$ super Yang-Mills by means of D5-branes wrapped 
on a resolved Calabi-Yau space. This is done by fixing an ambiguity on 
the correct interpretation of the holographic dual of the running 
gauge coupling and amounts to identify a specific 2-cycle in the dual 
ten-dimensional supergravity background. In doing so, we also discuss 
the role played in this context by gauge transformations in the 
relevant seven-dimensional gauged supergravity. While all nice 
properties of the duality are maintained, this modification of the 
dictionary has some interesting physical consequences and solves a 
puzzle recently raised in the literature. In this refined framework, 
it is also straightforward to see how the correspondence naturally 
realizes a geometric transition. 
\end{abstract} 
\end{titlepage} 
 
 
 
 
\section{Introduction and summary of the results} 
One of the goals recently pursued in the context of the AdS/CFT 
correspondence has been to look for gravity duals of ${\cal N}=2$ and 
${\cal N}=1$ super Yang-Mills (SYM) theory, both with and without 
matter. 
 
An important step toward this goal was done by Maldacena and Nu\~nez 
(MN) in Ref.~\cite{Maldacena:2000yy} where a supergravity dual of pure 
${\cal N}=1$ SYM was proposed. As it is the case for gravity duals of 
confining gauge theories, one cannot obtain an exact duality since 
extra degrees of freedom, not belonging to the gauge theory, cannot be 
decoupled within the supergravity regime\footnote{Similar 
considerations hold, for instance, for another notable example of a 
gravity  dual of ${\cal N}=1$ SYM, the Klebanov-Strassler solution 
\cite{Klebanov:2000hb}, which, although displaying a different UV 
completion, is equivalent to the MN solution in the 
IR.}. Nevertheless, many interesting properties of the gauge theory 
are encoded in the dual supergravity background and can be described 
in detail. 
 
The MN model is constructed engineering a ${\cal N}=1$ SYM theory by
wrapping $N$ D5-branes on a non-trivial 2-cycle of a resolved
Calabi-Yau (CY) space. The unwrapped part of the brane world-volume
remains flat and supports a four-dimensional gauge theory. By
implementing the proper topological twist so to preserve 4
supercharges \cite{Bershadsky:1996qy}, some of the world-volume fields
become massive and decouple and one ends up, in the IR, with
four-dimensional pure ${\cal N}=1$ SYM theory.  This is obtained by
considering the world-volume theory of the D5-branes at energies where
both the higher string modes as well as the KK excitations on the
2-cycle decouple. The back-reaction of the D-branes deforms the
original background. The topology of the resulting space is in general
very different from the starting CY space. In this case, as discussed
by Vafa in Ref.~\cite{Vafa:2000wi}, one expects the resulting space to
be a deformed CY space, where the 2-cycle has shrunk but a 3-sphere has
blown-up, rendering a ten-dimensional non-singular solution. The
question is whether one can extract information on the gauge theory,
possibly at non-perturbative level, from the dual supergravity
background.
 
This question was recently addressed in a rather detailed way by Di
Vecchia, Lerda and Merlatti (DLM) in Ref.~\cite{DiVecchia:2002ks} (see
Ref.s~\cite{Loewy:2001pq,Apreda:2001qb} for previous works  discussing
these issues)  and a  number of informations on the gauge theory were
shown to be predicted  by the dual supergravity background in a
precise and quantitative  way. In particular, the expected running of
the gauge coupling with  the corresponding $\beta$-function, the
chiral symmetry anomaly,  the  phenomenon of gaugino condensation with
the corresponding breaking of  the chiral symmetry to $\Z_2$ in the IR
as well as the instanton  action contribution, were all derived from
the supergravity  solution. 

The gauge/gravity dictionary can be derived from two basic  equations
\cite{DiVecchia:2002ks} expressing the gauge coupling constant  and
the gaugino  condensate $\langle \lambda^2 \rangle $, which is a
protected operator  of the gauge theory, in terms of supergravity
degrees of  freedom. These two equations read
\begin{eqnarray} 
\label{fr} 
\frac{1}{g_{\rm YM}^2} &=& F(\rho) \sim {\rm Vol}(S^2) \\ 
\label{ar} \langle \lambda^2 \rangle &\sim 
&\left(\frac{\Lambda}{\mu}\right)^3 = G(\rho) 
\end{eqnarray} 
where $\Lambda$ is the dynamically generated scale, $\mu$ is the 
subtraction energy at which the gauge theory is defined and $F(\rho)$ 
and $G(\rho)$ are two given functions of the radial coordinate $\rho$ 
of the ten-dimensional supergravity background.  In particular, 
$F(\rho)$ is proportional to the volume of the 2-cycle the D5-branes 
wrap as seen in the deformed geometry. The identification in 
eq.\eqref{ar} (which is written in units of the energy scale) gives 
instead the radius/energy relation in the correspondence. 
 
An important point to notice is that the gravity quantities to be
compared with gauge theory operators should all be computed in the
ten-dimensional framework, this being the natural one from a string
theory point of view.  This was done only partially in
Ref.~\cite{Maldacena:2000yy}. The 2-cycle entering eq.~\eqref{fr} was
identified within the seven-dimensional gauged supergravity geometry,
while all other quantities, as the chiral anomaly and the gauge theory
instanton contribution were obtained considering the ten-dimensional
geometry. The observation above overcomes this hybrid interpretation
and leads to a very homogeneous picture of the entire duality, as it
was drawn in Ref~\cite{DiVecchia:2002ks}. However, as pointed out
recently in Ref.~\cite{Olesen:2002nh}, this posed a new problem since
it seemed that in doing so a singular transformation in the gauge
coupling was needed in order to get the NSVZ $\beta$-function
\cite{Novikov:1983uc} from the $\beta$-function obtained from the
corresponding gravitational dual.
 
In this paper we reconsider this issue, and clarify what is the
correct 2-cycle in the ten-dimensional deformed geometry to be
considered, related to the 2-cycle of the original resolved CY space
used to engineer the ${\cal N}=1$ SYM theory. It turns out that the
2-cycle considered in the literature is not the correct one. As we are
going to show, this observation solves the problem raised in
Ref.~\cite{Olesen:2002nh}, without spoiling, on the other hand, all
nice results obtained in Ref.~\cite{DiVecchia:2002ks}. In particular,
we will get their same result for the $\beta$-function, but
determining now unambiguously the two-loop coefficient. It turns out
that supergravity, through the holographic relations \eqref{fr} and
\eqref{ar}, gives a $\beta$-function which is in the same scheme as
that obtained by NSVZ, the Pauli-Villars scheme. Redefinitions of the
holographic relation \eqref{ar} by means of analytic functions of the
gauge coupling respecting the symmetry of $\langle\lambda^2 \rangle$,
correspond to a change of regularization scheme. This modifies the
$\beta$-function beyond two loops only, showing that supergravity
naturally respects the expected universality of the 2-loop
coefficient\footnote{We thank Wolfang Mueck for sharing with us  his
recent findings on related topics.}. As a further check for the
validity of our analysis, in this refined framework  it is easy to see
that the MN model realizes a geometric transition, as predicted for
these kind of gauge/gravity dualities by the general picture discussed
by Vafa in Ref.~\cite{Vafa:2000wi}.

\section{The geometry revisited} 
Let us start by summarizing the explicit form of the MN solution.
This solution is obtained from a non-singular domain wall solution of
seven-dimensional gauged supergravity \cite{cv}, parameterized by
coordinates $(x_0,\ldots,x_3,\rho,\theta_1,\phi_1)$, uplifting to
ten dimensions along a 3-sphere
\cite{Cvetic:2000dm,Chamseddine:1999uy}, parameterized by coordinates
$(\psi,\theta_2,\phi_2)$. The relevant fields (the metric, the dilaton
and the RR 3-form the D5-branes magnetically couple to) are
\begin{eqnarray} 
\label{mnsol} ds^2 &=& e^\Phi dx^2_{1,3} + e^{\Phi} \alpha' g_s N 
\left[ e^{2h} \left( d\theta_1^2 + \sin^2 \theta_1 \,d\phi_1^2 \right)
+ d\rho^2 + \sum_{a=1}^3\left(\sigma^a - A^a \right)^2 \right] \\
\label{mnsol1} e^{2 \Phi} &=& \frac{\sinh 2\rho}{2\, e^{h}} 
\\
\label{mnf3} F^{(3)} &=& 2 \, \alpha' g_s N \, \prod_{a=1}^3 
\left( \sigma^a - A^a \right) - \alpha' g_s N \, \sum_{a=1}^3 F^a
\wedge \sigma^a
\end{eqnarray} 
where
\begin{eqnarray} 
\label{mnsol2} A^1 &=& - \frac{1}{2} \,a(\rho) \,d\theta_1 \spa A^2 = 
\frac{1}{2} \,a(\rho) \,\sin\theta_1 \,d \phi_1 \spa A^3 = -
\frac{1}{2} \,\cos \theta_1 \,d\phi_1 \\
\label{mnsol3} e^{2h} &=&  \rho \coth 2 \rho - 
\frac{\rho^2}{\sinh^2 2 \rho} -\frac{1}{4} \spa a(\rho) = \frac{2
\rho}{\sinh 2 \rho}
\end{eqnarray} 
$A^a$ being the three $SU(2)_L$ gauge fields of the relevant
seven-dimensional gauged supergravity. The $\sigma^a$ are the
left-invariant one-forms parameterizing the 3-sphere
\begin{eqnarray} 
\label{sigmas} 
\sigma^1 &=& \frac{1}{2} \left(\cos \psi \,d\theta_2 + \sin \psi
\sin\theta_2 d\phi_2\right) \spa  \sigma^2 = - \frac{1}{2} \left(\sin
\psi \,d\theta_2 - \cos \psi \sin\theta_2 d\phi_2\right) \nonumber \\
\sigma^3 &=& \frac{1}{2} \left(d\psi + \cos\theta_2 d\phi_2 \right)
\end{eqnarray} 
Let us now turn to the identification of the actual $S^2$ of the
ten-dimensional geometry \eqref{mnsol} entering the gauge/gravity
relation \eqref{fr}. Naively one would say that this cycle is the
cycle parameterized by the two coordinates ($\theta_1 , \phi_1$).
This is indeed the original cycle already present in the
seven-dimensional solution one starts from to derive the
ten-dimensional one. This was the choice made both in
Ref.~\cite{Maldacena:2000yy} and Ref.~\cite{DiVecchia:2002ks}, within
the seven and ten-dimensional geometry, respectively. In fact, the
seven-dimensional solution is non trivially embedded in ten
dimensions, the non triviality coming from the topological twist 
performed in seven dimensions. As a result of this, there is a
non-trivial mixing between the three coordinates of the $S^3$ along
which one uplifts the solution ($\theta_2 , \phi_2 , \psi$) and those
of the $S^2$ along which the original seven-dimensional domain wall is
wrapped ($\theta_1 , \phi_1$). This mix can be seen explicitly by the
appearance of the seven dimensional gauge connection in the ten
dimensional metric \eqref{mnsol}. We could say that the
seven-dimensional domain wall already knows about the ten-dimensional
geometry via the twist, that from a seven-dimensional point of view
actually mixes space-time degrees of freedom with internal ones (note
that in ten dimensions all these degrees of freedom are relative to
space-time). For this reason, it will turn out that the proper 2-cycle
is different from that suggested by the naive intuition.
 
To identify the relevant 2-cycle (and the 3-cycle dual to it) we now
focus on the five-dimensional angular part of the metric
(\ref{mnsol}). Let us consider two particular limits, $\rho\to\infty$
and $\rho\to 0$. At large $\rho$, from the solution \eqref{mnsol} we
easily get
\begin{equation} 
ds^2_5 ~ \sim~ \rho\, (d\theta_1^2 + \sin ^2\theta_1 d\phi_1^2) +
\frac{1}{4}   (d\theta_2^2+\sin^2\theta_2 d\phi_2^2) +\frac{1}{4}
(d\psi + \cos\theta_1 d\phi_1 + \cos\theta_2 d\phi_2)^2
\end{equation} 
It is easy to see that this is precisely the metric of the $T^{1,1}$
manifold, that topologically is $S^2\times S^3$. Even if it differs
from the 'standard' $T^{1,1}$ (see for instance
Ref.s~\cite{Minasian:1999tt,Brandhuber:2001yi}) as now it is re-scaled
in a way it is no longer an Einstein space, we can anyhow determine
the non-trivial cycles. They are those of the standard $T^{1,1}$,
since the only difference with the above manifold is just a metric
difference.
 
In the above set of coordinates, see
Ref.s~\cite{Gubser:1998fp,Dasgupta:1999wx,Papadopoulos:2000gj,Minasian:2001sq},
the 2-cycle is not uniquely defined and it turns out there are two
different, but physically equivalent, choices
\begin{eqnarray} 
\label{ciclo0} 
S^2 &:& \theta_1=-\theta_2 \spa \phi_1=-\phi_2 \spa \psi=0\\
\label{ciclop} S^2 &:& \theta_1=\,\theta_2 \;\,\spa \phi_1=-\phi_2 
\spa \psi=\pi
\end{eqnarray} 
The value of $\psi$ is fixed by the physical requirement that the
cycle is that of minimal volume, this being proportional to the
wrapped D5 brane tension. It is easy to show that with the first
choice, $\theta_1=-\theta_2 \,, \,\phi_1=-\phi_2$, the minimal volume
in the geometry \eqref{mnsol} is for $\psi=0$. Analogously, for
$\theta_1=\theta_2 \,, \,\phi_1=-\phi_2$ we have that $\psi=\pi$.  Let
us stress that the two 2-cycles \eqref{ciclo0} and \eqref{ciclop} are
physically equivalent. Indeed we can see from eq.s~\eqref{mnsol} and
\eqref{mnf3} that the two corresponding volumes are equal, namely
\begin{equation} 
\label{vol} \mbox{Vol}(S^2) ~\sim~ \left[{\rm 
e}^{2h(\rho)}+\frac{1}{4}(a(\rho)-1)^2~\right] (d\theta^2 +
\sin^2\theta d\phi^2)
\end{equation} 
and the projection of the RR field strength along both cycles vanishes
at the origin, as it should be. Moreover, all the gauge theory
implications we will discuss in the next section are the same for the
two cycles.
 
As already discussed the 3-cycle is instead parameterized by
\begin{eqnarray} 
\label{s3} S^3 \::\; \theta_1 = \phi_1=0 
\end{eqnarray} 
Let us now study the metric at the origin. It has the following form
\begin{eqnarray} 
&&ds^2_5 \sim \frac{1}{4}\,(\cos\psi\sin\theta_2 d\phi_2 -\sin\psi d\theta_2
-\sin\theta_1 d\phi_1)^2 +\nonumber\\
&&+\,\frac{1}{4}\,(\sin\psi\sin\theta_2 d\phi_2 +\cos\psi d\theta_2
+d\theta_1)^2  +\frac{1}{4} (d\psi +
\cos\theta_1d\phi_1+\cos\theta_2d\phi_2)^2
\end{eqnarray} 
This is precisely the metric of a deformed conifold at the apex, see
Ref.~\cite{Minasian:1999tt}\footnote{We thank E. Gimon for a useful comment on this point.}. The
parameterization of the non trivial 2 and 3-cycle is known for this
metric, and is consistent with the ones found before. By implementing
eq.~\eqref{ciclo0} (or equivalently eq.~\eqref{ciclop}) and
eq.~\eqref{s3} in the above metric one finds a vanishing radius for
the 2-sphere and a finite one for the 3-sphere, as expected for a
deformed conifold. We will come back to this issue in the last section.
 
Let us anticipate that with the above identification of the 2-cycle,
which has of course non-trivial consequences on the explicit form of
the function entering in eq.~\eqref{fr} (see the explicit expression
in eq.~\eqref{vol}), the main results about the gauge theory obtained
in Ref.~\cite{DiVecchia:2002ks} do not change drastically. On the
other hand, as already noticed, the problem related to the
determination of the proper $\beta$-function by means of the
gravitational dual will be solved.
 
Before studying the gauge theory implications of what we have been
discussing so far, we want to illustrate another way one can get the
same result. In doing so, we also clarify the meaning of the
seven-dimensional gauge transformations. Indeed in the non singular
seven dimensional solution we are free to make $SU(2)$ gauge
transformations
$$ A \to g^{-1}A \, g + {\rm i} \, g^{-1}d g $$ where $g$ is an
element of the $SU(2)$ group and $A$ is the $SU(2)$ gauge connection.
 
The ten-dimensional solution then is not completely determined, even
if all the possible solutions should be equivalent. Indeed different
seven dimensional gauge choices correspond to different
parameterizations of the relevant ten dimensional geometry.  We show
this with one concrete example. Consider then the following gauge
transformation
\begin{equation} 
g ~=~ {\rm e}^{- \frac{\rm i}{2}\theta_1\sigma_1} {\rm e}^{- \frac{\rm
i}{2} \phi_1\sigma_3}
\end{equation} 
on the $A^a$'s in eq.~\eqref{mnsol2}. The new gauge connection is
\begin{eqnarray} 
A'^1 &=& \frac{1}{2}\,(a(\rho) - 1)\, (-\cos\phi_1
d\theta_1~+~\cos\theta_1 \sin\theta_1 \sin\phi_1 d\phi_1) \nonumber \\
A'^2 &=& \frac{1}{2}\,(a(\rho) - 1)\, (\sin\phi_1
d\theta_1~+~\cos\theta_1\sin\theta_1 \cos\phi_1 d\phi_1)
\label{gauge} \\ 
A'^3 &=& -\frac{1}{2} \,(a(\rho) - 1)\, \sin^2\theta_1 d\phi_1
\nonumber
\end{eqnarray} 
where the function $a(\rho)$ is again given by eq.~\eqref{mnsol3}.
Now we have that $A'\to 0$ for $\rho\to 0$. As already discussed at
the beginning of this section, the seven-dimensional gauge connection
is the field responsible for the non-trivial mixing between the
seven-dimensional coordinates and the ten-dimensional ones. Moreover,
the seven-dimensional solution represents a domain wall located
precisely at $\rho=0$ (corresponding to the wrapped D5-branes). Hence
if the gauge potential vanishes at $\rho=0$, the 2-cycle no longer
mixes with the $S^3$ used to uplift the solution to ten
dimensions. Then, also in the ten-dimensional solution, the 2-cycle
will be simply parameterized by $\theta_1$ and $\phi_1$, while, as
usual, the 3-cycle by $\theta_2,\ \phi_2$ and $\psi$.
 
Once the cycles are properly identified, it is completely equivalent
to study the gauge theory by means of this solution (that in terms of
$A'$, eq.~\eqref{gauge}) or of the other one (that in terms of $A$,
eq.~\eqref{mnsol2}). All the physical results  we are going to
describe in the next section will not change. Note that also for this
cycle the $\rho$-dependent volume is precisely given by
eq.~\eqref{vol} and the projection of the RR field-strength on it
vanishes at the origin.

\section{The gauge/gravity dictionary revisited} 
 
Let us now investigate what are the consequences of the above
discussion on the gauge/gravity dictionary. As recalled in the
introduction, the two crucial equations in relating gauge and gravity
quantities are those expressing the gauge coupling and the energy
scale of the gauge theory as functions of gravity fields.  Using the
solution \eqref{mnsol}-\eqref{mnsol3} they read in our case
\begin{eqnarray} 
\label{gym1} \frac{1}{g_{\rm YM}^2} &=& 
\frac{1}{2(2\pi)^3 \alpha'g_s} \int_{S^2}{\rm  e}^{-\phi}\sqrt{
\mbox{det}G} \,= \,\frac{N}{16 \pi^2} \,Y(\rho) \\
\label{cond1} \left(\frac{\Lambda}{\mu}\right)^3 &=& a(\rho) 
\end{eqnarray} 
where
\begin{equation} 
Y(\rho) = 4 \,e^{2 h(\rho)} + \left( a(\rho) -1\right)^2 = 4 \rho
\,\tanh \rho \spa a(\rho) = \frac{2\rho }{\sinh 2\rho}
\end{equation} 
Eq.~\eqref{gym1} is obtained identifying the Yang-Mills coupling
constant from the DBI action of the D5-branes while the energy/radius
relation \eqref{cond1} was obtained in Ref.~\cite{DiVecchia:2002ks}
from the identification of the gaugino condensate in terms of the
supergravity field $a(\rho)$ \cite{Apreda:2001qb}.

Eq.~\eqref{gym1} differs from the analogous equation of 
Ref.~\cite{DiVecchia:2002ks}, eq.~(4.7), the difference being in the 
precise $\rho$-dependence of the function $Y(\rho)$ (that is 
essentially the volume of the $S^2$). In particular, now $Y(\rho)$ 
goes to zero at small $\rho$, and we get 
\begin{eqnarray} 
\label{gyminf} 
\frac{1}{g_{\rm YM}^2} &\simeq & \frac{N\rho}{4 \pi^2} \qquad {\rm 
for} \quad \rho \rightarrow \infty \quad {\rm which\; means} \quad \mu 
>> \Lambda \\ \label{gym0} \frac{1}{g_{\rm YM}^2} &\simeq & \,\,0 
\;\;\; \qquad {\rm for} \quad  \rho \rightarrow 0 \quad \;\;{\rm 
which\; means} \quad \mu \sim \Lambda 
\end{eqnarray} 
The large $\rho$ behavior is the same as in DLM, while at $\rho=0$ we 
get a Landau pole. We will comment more on this point later. 
 
 From the above equations one can get the complete perturbative ${\cal 
N}=1$ $\beta$-function. We can write 
\begin{equation} 
\beta(g_{\rm YM}) = \frac{\partial g_{\rm YM}}{\partial \ln 
(\mu/\Lambda)} = \frac{\partial g_{\rm YM}}{\partial 
\rho}\frac{\partial \rho}{\partial \ln (\mu/\Lambda)} 
\end{equation} 
and compute the two derivative contributions from eq.~\eqref{gym1} and 
eq.~\eqref{cond1}, respectively. In doing so, let us first disregard 
the exponential corrections, which are sub-leading at large $\rho$ and 
which give rise to non-perturbative contributions. In this case the 
expansion in eq.~\eqref{gyminf} is exact. We easily get 
\begin{equation} 
\label{rolog1} \frac{\partial g_{\rm YM}}{\partial \rho} = - 
\frac{N g_{\rm YM}^3}{ 8 \pi^2} \spa \frac{\partial \rho}{\partial \ln 
(\mu/\Lambda)} = \frac{3}{2} \, \left(1 - \frac{1}{2 \rho}\right)^{-1} 
= \frac{3}{2} \, \left(1 - \frac{N g_{\rm YM}^2}{8 \pi^2}\right)^{-1} 
\end{equation} 
where in the last step of the second equation we have used again 
eq.~\eqref{gyminf}. The final result is then 
\begin{equation} 
\label{betapert1} 
\beta(g_{\rm YM}) = - 3 \, \frac{N g_{\rm YM}^3}{16 \pi^2}  \left(1 - 
\frac{N g_{\rm YM}^2}{8 \pi^2} \right)^{-1} 
\end{equation} 
which is the NSVZ $\beta$-function \cite{Novikov:1983uc}. Note how 
this differs from the result of DLM. Besides exponentially suppressed 
corrections, in their case the expression \eqref{gyminf} received also 
sub-leading corrections as power series in $1/\rho$ and $\log 
\rho$. These corrections should be taken into account when deriving 
the perturbative $\beta$-function. The contributions in the $1/\rho$ 
change the result beyond two loop only, hence respecting the 
universality of the two-loop coefficient of the $\beta$-function. The 
contributions proportional to $\log \rho$, instead, spoil this 
universality. This gives, as a result, a $\beta$-function not 
belonging to the same universality class of the NSVZ 
$\beta$-function. As discussed in Ref.~\cite{Olesen:2002nh}, in order 
to get rid of the unwanted logarithmic corrections and get a 
$\beta$-function respecting the universality of the two loop 
coefficient, a singular transformation in the gauge coupling is 
needed. We have shown here that the correct identification of the 
relevant 2-cycle in the geometry gives instead directly the result 
\eqref{betapert1} and the  complications discussed in 
Ref.~\cite{Olesen:2002nh} are not present. Let us stress that this is 
not an option: once the correct 2-cycle is identified, the result 
\eqref{betapert1} naturally follows. 
 
Note also how the correct gauge/gravity dictionary naturally 
respects the universality of the two-loop coefficient of the 
$\beta$-function. Indeed the geometric considerations leading to the 
identification of the gaugino condensate with the function $a(\rho)$ 
\cite{Apreda:2001qb,DiVecchia:2002ks} are insensible to a redefinition 
of the holographic relation \eqref{cond1} by means of an analytic 
function of the gauge coupling \cite{Olesen:2002nh}.  If doing so, one 
can easily see that the result we have obtained, 
eq.~\eqref{betapert1}, changes beyond two loops only. 
 
As anticipated there are also some non-perturbative contributions to 
the $\beta$-function that supergravity suggests should be 
present. These are included by considering the full expression for 
$Y(\rho)$ and $a(\rho)$ in eq.s~\eqref{gym1} and \eqref{cond1}.  The 
analysis performed in Ref.~\cite{DiVecchia:2002ks} is essentially 
unchanged in this case and we do not repeat it here.  It would be nice 
to check this (unexpected) prediction by doing some computations in 
the field theory. 
 
The correct supergravity prediction for the chiral anomaly and chiral 
symmetry breaking discussed in Ref.~\cite{DiVecchia:2002ks} is also 
unchanged. The gauge theory $\theta$-angle is related to the flux of 
the RR 2-form $C^{(2)}$ through the 2-cycle and the $N$ vacua of the 
gauge theory are parameterized by shifts in the angular variable 
$\psi$. Finally, the gauge theory instantons are described by 
euclidean D1-branes wrapped on the 2-cycle \eqref{ciclo0} (or 
equivalently \eqref{ciclop}), and computing their corresponding action 
in the background \eqref{mnsol}-\eqref{mnsol3} one easily finds the 
expected gauge theory  instanton action, as in 
Ref.~\cite{DiVecchia:2002ks}. 
 
Summarizing, once the proper identification of the $S^2$ related to 
the gauge coupling is made, all the nice properties of the 
correspondence discussed in Ref.~\cite{DiVecchia:2002ks} still hold 
while the complications addressed in Ref.~\cite{Olesen:2002nh} turn 
out not to be present. The only property which is lost is soft 
confinement one had signs of, in the DLM picture, when taking $\rho$ 
all the way to zero. We find a Landau pole, instead. However, this is 
not really an issue. The curvature of the MN background goes like 
$\alpha'{\cal R} \sim 1/g_s N$ so the regime in which the supergravity 
approximation is reliable is for large $N$. In this regime a Landau 
pole can indeed be present even if the gauge coupling remains finite 
at the scale $\Lambda$, since in eq.~\eqref{gym0} it is really $g_{\rm 
YM}^2 N$ which is going to infinity and not the gauge coupling 
itself. To discuss the duality in the deep IR at finite $N$, one has 
to go beyond the supergravity approximation.

\section{The duality as a geometric transition} 
 
 
As anticipated, a by-product of our analysis is that now it is easy to 
show that the MN solution is indeed an explicit example realizing the 
general picture proposed by Vafa in Ref.~\cite{Vafa:2000wi} (see 
Ref.s~\cite{Cachazo:2001jy,Dasgupta:2001ac} for further 
clarifications). 
 
The general idea discussed in Ref.~\cite{Vafa:2000wi}, applied to the 
case at hand, is to engineer a supersymmetric gauge theory by means of 
D5-branes wrapped on a supersymmetric 2-cycle of a resolved CY 
manifold. The dual supergravity solution is conjectured to correspond 
to a deformed CY geometry, where the D-branes are absent and the 
manifold has undergone a geometric transition: on the deformed CY the 
$S^2$ is shrunk and an $S^3$ has blown-up. The D-branes are replaced 
by $H_{NSNS}$ flux through a non-compact 3-cycle and $H_{RR}$ flux 
through the $S^3$. 
 
From the discussion in section 2, it is clear that the MN duality 
indeed realizes a geometric transition. Starting from the 
ten-dimensional metric, eq.~\eqref{mnsol}, and taking the limit $\rho 
\to 0$ we get 
\begin{equation} 
\label{mn0} ds^2 \sim  \, dx^2_{1,3} + \alpha' g_s N\, \left[ 
d\rho^2 + \rho^2 \left( d\theta_1^2 + \sin^2\theta_1 \,d\phi_1^2 
\right) + \sum_{a=1}^3\left(\sigma^a - A^a \right)^2 \right] 
\end{equation} 
By using eq.s~\eqref{s3}, combined with one parameterization of the 
2-cycle (equivalently eq.s~\eqref{ciclo0} or \eqref{ciclop}) one 
immediately sees that the topology of the space at $\rho=0$ is that of 
an $S^3$, which is blown-up: while the 2-sphere is shrunk ($R^2_{S^2} 
\sim \rho^2$), the radius of the 3-sphere remains finite, $R^2_{S^3} = 
\alpha' g_s N$. Hence the original resolved CY space used to engineer 
the ${\cal N}=1$ SYM by means of D-branes wrapped on a non-vanishing 
2-cycle has undergone a geometric transition to a deformed CY, where 
the $S^2$ has shrunk and an $S^3$ has blown-up, as predicted by Vafa 
duality\footnote{The same conclusion could be reached considering the 
background defined by the seven dimensional gauge connection 
\eqref{gauge}.}. 
 
Let us end noticing an aspect where the MN correspondence is 
apparently different from Vafa general picture. In the MN supergravity 
solution there is just one 3-form, $H_{RR}$, switched-on while the 
NS-NS one is not. As we have been extensively discussed, in the MN 
solution the gauge coupling is related to the volume of the $S^2$, 
rather than to the $B_{NS}$ flux along the 2-cycle, as it is instead 
the case for Vafa duality. In fact, the MN configuration is related by 
T-dualities to fractional D3-branes on ${\cal N}=1$ orbifolds (having 
D4 branes suspended between non-parallel NS5 branes as an intermediate 
step). There, the volume of $S^2$ translates indeed into the $B_2$ 
flux along the $S^2$. So, in a sense, this difference amounts just to 
a U-duality gauge.

\section*{Acknowledgments} 
We are grateful to P. Di Vecchia and A. Lerda for some very useful
remarks on a preliminary version of this paper. We thank
D. Berenstein, C. Nu\~nez and E. Imeroni for discussions and/or
correspondence. M.B. is supported by
an EC Marie Curie Postdoc  Fellowship under contract number
HPMF-CT-2000-00847. Work partially  supported by the European
Community's Human Potential Program under  contract number
HPRN-CT-2000-00131.

\end{document}